\documentclass[twocolumn,prb,floatfix]{revtex4-2}
\usepackage{amsfonts}
\usepackage{amsmath,amsbsy,amssymb}
\usepackage[dvipdfmx]{graphics}
\usepackage{float}
\usepackage{bm}
\usepackage{braket}
\usepackage{mathtools}
\usepackage[breaklinks,colorlinks=true,linkcolor=blue,urlcolor=blue,citecolor=blue]{hyperref}
\usepackage{verbatim}

\usepackage{color,xcolor}
\usepackage[normalem]{ulem}

\newcommand{\1}{\mbox{1}\hspace{-0.25em}\mbox{l}}

\begin{document}

\title{Discriminant indicator with generalized rotational symmetry}
\author{Hiromasa Wakao$^1$}
\author{Tsuneya Yoshida$^2$}
\author{Yasuhiro Hatsugai$^2$}%

\affiliation{%
$^1$Graduate School of Pure and Applied Sciences, University of Tsukuba, Tsukuba, Ibaraki 305-8571, Japan\\
$^2$Department of Physics, University of Tsukuba, Tsukuba, Ibaraki 305-8571, Japan
}%
\date{\today}

\renewcommand{\figurename}{Fig.}

%Abstract
\begin{abstract}
 Discriminant indicators with generalized inversion symmetry
 are computed only from data at the high-symmetry points.
 They allow a systematic search for exceptional points.
 In this paper, we propose discriminant indicators for two- and three-dimensional systems with generalized $n$-fold rotational symmetry ($n=4$, $6$).
 As is the case for generalized inversion symmetry, the indicator taking a nontrivial value predicts the emergence of exceptional points and loops without ambiguity of the reference energy.
 A distinct difference from the case of generalized inversion symmetry is that the indicator with generalized $n$-fold rotational symmetry ($n=4$, $6$)
 can be computed only from data at two of four high-symmetry points in the two-dimensional Brillouin zone.
 Such a difference is also observed in three-dimensional systems.
 Furthermore, we also propose how to fabricate a two-dimensional system with generalized four-fold rotational symmetry for an electrical circuit.
\end{abstract}

\maketitle
%Introduction
\section{Introduction\label{sec:introduction}}
In recent decades, many efforts have been devoted to understanding the topological properties of wavefunctions~\cite{Hatsugai1993_2,Hatsugai1993,Kane2005,Teo2008,Qi2008,Hasan2010,Qi2011,Sato2016}.
In particular, it turns out that symmetry enriches topological structure~\cite{Chiu2016},
which is exemplified by $\mathbb{Z}_2$ topological insulators with time-reversal symmetry~\cite{Fu2006,Fu2007_2,Moore2007,Zhang2009,Roy2009,Hsieh2009}
and topological crystalline materials~\cite{Zak1989,Fu2007,Hughes2011,Fu2011,Fang2012,Slager2013,Morimoto2013,Koshino2014,Kim2015}.
Intriguingly, crystalline symmetry also plays a key role in determining topological properties.
For instance, in the presence of the inversion symmetry and time-reversal symmetry,
the topological invariant can be computed only from the parity eigenvalues at high-symmetry points in the Brillouin zone (BZ)~\cite{Fu2007}.
This notion is generalized in Refs.~\cite{Po2017,Kruthoff2017,Watanabe2018,Ono2018,Po2020}
which introduced symmetry indicators
as powerful tools for the systematic search for topological materials.

In parallel with this progress, recent extensive studies have opened up a new arena of topological physics:
non-Hermitian systems~\cite{Hatano1996,Bender1998,Rotter2009,Sato2012,Ashida2016,Ashida2017,Bergholtz2021,Ashida2020}.
In these systems, the eigenvalues of the Hamiltonian may become complex,
which induces exotic phenomena~\cite{Esaki2011,Hu2011,Liang2013,Gong2017,Shen2018,Carlstrom2018,Liu2019_2,Carlstrom2019,Okuma2019,Yoshida2019_2,Gong2018,Kawabata2019_2,Zhou2019,Chen2022}
such as the emergence of exceptional points~\cite{Zhen2015,Lee2016,Leykam2017,Kozzi2017,Alvarez2018,Yoshida2018,Kawabata2019,Wojcik2020,Yang2021,Rui2022}
and skin effects~\cite{Kunst2018_2,Yao2018,Yao2018_2,Edvardsson2019,Lee2019,Borgina2020,Zhang2020,Okuma2020,Okugawa2020,Okuma2021,Sun2021}.
At the exceptional points, non-Hermitian topological properties protect band touching for both of the real and imaginary parts of the eigenvalues
which are further enriched by symmetry~\cite{Budich2019,Okugawa2019,Yoshida2019_3,Zhou2019_2,Mandal2021,Delplace2021}.
The skin effects result in anomalous sensitivity of energy spectrum on the presence or absence of boundaries~\cite{Kunst2018_2,Yao2018,Yao2018_2,Lee2019,Edvardsson2019,Borgina2020,Zhang2020,Okuma2020}.
The exceptional points and skin effects have been reported for a wide range of systems such as
open quantum systems~\cite{Jin2009,Jose2016,Lee2016,Xu2017,Yoshida2020_4}, electrical circuits~\cite{Ezawa2019,Hofmann2019,Ezawa2019_2,Yoshida2020,Hofmann2020,Helbig2020}, phononic systems~\cite{Miri2019,Yoshida2019,Ghatak2020,Scheibner2020}, photonic crystals~\cite{Guo2009,Ruter2010,Feng2017,Isobe2021}, and so on.

For the systematic search for non-Hermitian topological systems, symmetry indicators are powerful tools.
Indeed, recent works have introduced indicators for non-Hermitian systems~\cite{Okugawa2020,Yoshida2020_4,Okugawa2021,Vecsei2021,Shiozaki2021,Tsubota2022}.
Among them, the discriminant indicator~\cite{Yoshida2022} captures exceptional points without ambiguity of the reference energy.

Despite its usefulness,
the discriminant indicator is introduced only for systems with generalized inversion symmetry~\cite{Yoshida2022}.
For the systematic search for exceptional points,
the extension to other symmetries is crucial.

In this paper,
we extend the indicator to systems with generalized $n$-fold
rotational ($C_n$) symmetry for $n=4$, $6$.
The indicators successfully capture the exceptional points in two dimensions
without ambiguity of the reference energy,
which is demonstrated by systematic analysis of toy models.
We also introduce indicators which capture exceptional loops in the three-dimensional BZ.
In contrast to the case with the generalized inversion symmetry~\cite{Yoshida2022},
the indicator with generalized $n$-fold rotational symmetry ($n=4$, $6$) can be computed only from data at two of the four high-symmetry points
in the two-dimensional BZ.
Such a difference is also observed in three-dimensional systems.
We also show that the nontrivial value of the discriminant indicator with generalized $C_4$ symmetry
in three dimensions predicts the merging of the exceptional loops at the line specified by $(k_x,k_y)= (0,0)$ or $(\pi, \pi)$.
Moreover, we propose an experimental realization of the topoelectrical system with
generalized $C_4$ symmetry.

%Organization of this paper
The rest of this paper is organized as follows.
In Sec.~\ref{sec:discr-indic-with},
we derive the indicators for systems with generalized $C_n$ symmetry ($n=4$, $6$).
In Sec.~\ref{sec:toy-model-with},
we demonstrate that these indicators capture exceptional points and loops
by numerically analyzing toy models with generalized $C_n$ symmetry ($n=4$, $6$).
In Sec.~\ref{sec:topo-syst-with},
we experimentally realize the topoelectrical circuit with generalized $C_4$ symmetry.
In Sec.~\ref{sec:summary}, we present a summary of this paper.

\begin{figure*}[t]
 \centering
\includegraphics[trim={0cm 0cm 0cm 0cm},width =\hsize]{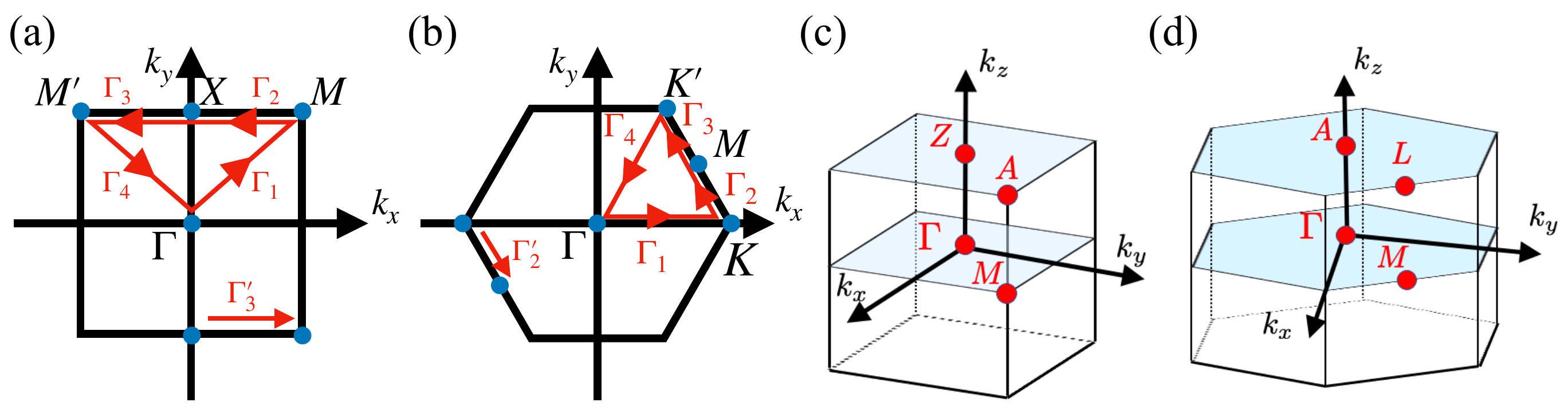}
\caption{(Color online)
 (a) and (b) Sketches of the BZ and the path of integrations in Eq.~(\ref{eq:13}).
 (a) The BZ for systems with generalized $C_4$ symmetry.
 The high-symmetry points $\Gamma$, $M$, $Y$, and $M^{\prime}$ are denoted by blue dots.
 The paths $\Gamma_1$, $\Gamma_2$, $\Gamma_3$, and $\Gamma_4$ connect high-symmetry points.
 The path $\Gamma_3^{\prime}$ is the mapped path of $\Gamma_3$
 after applying the generalized $C_4$ operator twice.
 (b)
 The BZ for systems with generalized $C_6$ symmetry.
 The high-symmetry points $\Gamma$, $K$, $M$, and $K^{\prime}$ are denoted by blue dots.
 The paths $\Gamma_1$, $\Gamma_2$, $\Gamma_3$, and $\Gamma_4$ connect high-symmetry points.
 The path $\Gamma_2^{\prime}$ is the mapped path of $\Gamma_2$
 after applying the generalized $C_6$ operator three times.
 (c) and (d) Schematic figures of the three-dimensional BZs.
 Blue planes denote the area for $k_z = 0$ and $\pi$.
 (c) The BZ of the tetragonal lattice.
 The high-symmetry points, $\Gamma$, $M$, $Z$, and $A$ are denoted by red dots.
 (d) The BZ of the hexagonal lattice.
 The high-symmetry points, $\Gamma$, $M$, $A$, and $L$ are denoted by red dots.
 }
 \label{fig:BZ_GRS}
\end{figure*}

\section{Discriminant indicator with generalized rotational symmetry\label{sec:discr-indic-with}}
In this section, we derive indicators for systems with generalized rotational symmetry.
In Sec.~\ref{sec:indic-gener-rotat}, along with a definition of generalized inversion symmetry,
we provide an overview of our main results.
The detailed derivation is provided in Secs.~\ref{sec:constr-discr-under}-\ref{sec:two-dimens-syst-1}.
In Sec.~\ref{sec:three-dimens-syst}, we also discuss indicators for exceptional loops in three-dimensional systems.

\subsection{Overview\label{sec:indic-gener-rotat}}
Consider an $N \times N$-Hamiltonian $H(\bm{k})$ for two-dimensional systems
with generalized $C_n$ symmetry under the periodic boundary condition~\cite{Okugawa2020,Vecsei2021,Sun2021,Shiozaki2021,Okugawa2021,Yoshida2022,Chen2022,Tsubota2022,Rui2022}.
In this case, the Hamiltonian satisfies
\begin{equation}
 \label{eq:6}
  U_{C_n} H(\bm{k}) U_{C_n}^{-1} = H^{\dagger} (R_n\bm{k}),
\end{equation}
with
\begin{equation}
 \label{eq:7}
  R_n =
  \begin{pmatrix}
   \cos 2\pi/n &  - \sin 2\pi/n \\
   \sin 2\pi/n & \cos 2\pi/n
  \end{pmatrix},
\end{equation}
and a unitary operator $U_{C_n}$ satisfying $U_{C_n}^n = \1_{N\times N}$.

The above non-Hermitian Hamiltonian may host exceptional points which are characterized by the discriminant number~\cite{Wojcik2020,Yang2021,Delplace2021}
\begin{equation}
 \label{eq:2}
  \nu = \oint \frac{d\bm{k}}{2\pi i} \cdot \bm{\nabla}_{\bm{k}} \ln \Delta (\bm{k}).
\end{equation}
The integral is taken along a closed path in the BZ.
The discriminant $\Delta (\bm{k})$ of the characteristic polynomial $P(\bm{k},E)=\det [ H(\bm{k})-E \1_{N \times N}] = \sum_{j=0}^N a_j E^j$ is defined as
\begin{equation}
 \label{eq:1}
     \Delta (\bm{k}) := \prod_{n> n^{\prime}} (\epsilon_n - \epsilon_{n^{\prime}})^2  .
\end{equation}
Here, $\epsilon_n$ denote the eigenvalues of $H(\bm{k})$.
$\Delta (\bm{k})$ can be rewritten as
\begin{widetext}
\begin{equation}
 \label{eq:10}
  \Delta (\bm{k})
  =
  \text{det}
   \begin{pmatrix}
   a_0    & a_1    & a_2    & \cdots & a_N     & 0      & \cdots  & 0      \\
   0      & a_0    & a_1    & \cdots & a_{N-1} & a_N    & \cdots  & 0      \\
   \vdots & \ddots & \ddots & \ddots & \vdots  & \ddots & \ddots  & \vdots \\
   0      & \cdots & a_0    & a_1    & a_2     & \cdots & a_{N-1} & a_N    \\
   b_1    & b_2    & b_3    & \cdots & b_N     & 0      & \cdots  & 0      \\
   0      & b_1    & b_2    & \cdots & b_{N-1} & b_N    & \cdots  & 0      \\
   \vdots & \ddots & \ddots & \ddots & \vdots  & \ddots & \ddots  & \vdots \\
   0      & \cdots & 0      & 0      & b_1     & b_2    & \cdots  & b_N
   \end{pmatrix},
\end{equation}
\end{widetext}
with $b_j = j a_j$~\cite{Mignotte1992,Lang2002}.
Clearly, $\Delta (\bm{k})$ is determined by the coefficients of the characteristic polynomial.
If $\nu$ takes a non-zero value,
$\Delta (\bm{k})$ vanishes at a point inside of the loop.
Such a point corresponds to an exceptional point
because Eq.~(\ref{eq:1}) indicates the band touching on this point.

In the presence of the generalized $C_n$ symmetry,
the discriminant $\Delta (\bm{k})$ satisfies
\begin{equation}
 \label{eq:45}
  \Delta (\bm{k}) = \Delta^{\ast} (R_n \bm{k}),
\end{equation}
which is proven in Sec.~\ref{sec:constr-discr-under}.

By making use of Eq.~(\ref{eq:45}), we obtain discriminant indicators for two-dimensional systems
with generalized $C_n$ symmetry ($n=4$, $6$),
\begin{subequations}
\begin{equation}
 \label{eq:43}
  (-1)^{\nu_{C_4}}: = \xi_{C_4,2D}   = \text{sgn} \Delta (\Gamma) \text{sgn} \Delta (M),
\end{equation}
\begin{equation}
 \label{eq:44}
  (-1)^{\nu_{C_6}}: = \xi_{C_6,2D} = \text{sgn} \Delta (\Gamma) \text{sgn} \Delta (M).
\end{equation}
\end{subequations}
Here, $\nu_{C_4}$ and $\nu_{C_6}$ denote the discriminant number
computed along the closed path illustrated in Fig.~\ref{fig:BZ_GRS}(a) [Fig.~\ref{fig:BZ_GRS}(b)].
Symbols $\Gamma$ and $M$ denote the high-symmetry points in the BZ [see Fig.~\ref{fig:BZ_GRS}].
Because of the constraint written in Eq.~(\ref{eq:45}),
the discriminant becomes real at $\Gamma$ and $M$ points.

We note that for $n=2$,
the problem is reduced to the case for the generalized inversion symmetry,
which is discussed in Appendix~\ref{sec:discr-numb-with}.
For systems with generalized $C_3$ symmetry, the Hamiltonian becomes
Hermitian~\footnote{
This can be seen as follows,
$H(\bm{k}) = U_{C_3}^3 H(\bm{k}) U_{C_3}^{-3} = H^{\dagger} (R_3^3 \bm{k}) = H^{\dagger}(\bm{k})$.}.
In Secs.~\ref{sec:two-dimens-syst} and \ref{sec:two-dimens-syst-1}, we derive these indicators.
These indicators can be computed without the input of the reference energy, and predict the presence of exceptional points.

\subsection{Constraints on the discriminant with generalized rotational symmetry\label{sec:constr-discr-under}}
For the systems with generalized $C_n$ symmetry, the polynomial $P(\bm{k},E)$ satisfies
\begin{equation}
\label{eq:8}
 \begin{split}
  P(\bm{k},E)
  &= \det [U_{C_n}^{\dagger} H^{\dagger} (R_{n} \bm{k}) U_{C_n} - E \1_{N\times N}]\\
  &= \det U_{C_n}^{\dagger} \det[H^{\dagger} (R_{n} \bm{k}) - E \1_{N\times N}] \det U_{C_n}\\
  &=  \det[H^{\ast} (R_{n} \bm{k}) - E \1_{N\times N}].
 \end{split}
\end{equation}
Here, we have used $\det A^T = \det A $ and $\det (AB) =\det A \det B = \det B \det A$.
We note that the discriminant can be computed only from coefficients of the polynomial.
It results in Eq.~(\ref{eq:45}).
It is straight forward to extend the above arguments to three-dimensional cases.
Specifically, for three-dimensional systems with generalized $C_n$ symmetry about the $z$axis,
the discriminant satisfies
\begin{equation}
 \label{eq:41}
 \Delta (\bm{k}_{\|},k_z) = \Delta^{\ast} (R_n \bm{k}_{\|},k_z),
\end{equation}
with $\bm{k}_{\|}^{T} = (k_x,k_y)$.

\subsection{Two-dimensional system with generalized fourfold rotational symmetry\label{sec:two-dimens-syst}}
In order to derive Eq.~(\ref{eq:43}), let us consider the discriminant number evaluated along the closed path illustrated in Fig.~\ref{fig:BZ_GRS}(a).
This integral can be decomposed into integrals along the path $\Gamma_i$ ($i=1, \cdots, 4$),
\begin{equation}
 \label{eq:13}
  p_{i} = \int_{\Gamma_i} \frac{d \bm{k}}{2\pi i} \cdot \bm{\nabla}_{\bm{k}} \ln \Delta (\bm{k}).
\end{equation}
Here, $\Gamma_i$ ($i=1,\cdots,4$) are the paths between the high-symmetry points [see Fig.~\ref{fig:BZ_GRS}(a)].
First,
we note that $p_2 + p_3$ vanishes
because applying the generalized $C_4$ operator twice maps
the path $\Gamma_3$ to the path $\Gamma_3^{\prime}$ which is equivalent to the path $\Gamma_2$ [see Fig.~\ref{fig:BZ_GRS}(a)]
due to the periodicity of the BZ.

Second, let us evaluate $p_1 + p_4$.
The integration of $\Delta (\bm{k})$ along the path $\Gamma_4$ is mapped to
that of $\Delta^{\ast} (\bm{k})$ along the opposite direction of $\Gamma_1$
by the generalized $C_4$ symmetry.
As a result, the integration $p_1 + p_4$ is simplified as
\begin{equation}
 \label{eq:14}
  \begin{split}
   2\pi i (p_1 + p_4)
   &= \int_{\Gamma_1}d \bm{k}  \cdot \bm{\nabla}_{\bm{k}} \ln \Delta ( \bm{k}) \\
   &
   + \int_{\Gamma_4}  d \bm{k} \cdot \bm{\nabla}_{\bm{k}} \ln \Delta^{\ast} (R_4^{-1} \bm{k}) \\
   & = \int_{\Gamma_1}  d \bm{k} \cdot \bm{\nabla}_{\bm{k}} \ln \Delta ( \bm{k})
   - \int_{\Gamma_1}  d \bm{k} \cdot \bm{\nabla}_{\bm{k}} \ln \Delta^{\ast} (\bm{k}) \\
   & = 2 i \text{Im}  \int_{\Gamma_1} d \bm{k} \cdot \bm{\nabla}_{\bm{k}}
  \ln \Delta ( \bm{k}) \\
      & = 2 i  \int_{\Gamma_1}  d \bm{k} \cdot \bm{\nabla}_{\bm{k}}
  \arg \Delta ( \bm{k}) \\
      & =2i \left[\arg \Delta (M ) - \arg \Delta (\Gamma)+2\pi N_0\right],
  \end{split}
\end{equation}
with some integer $N_0$.
From the second to the third lines, we have used Eq.~(\ref{eq:45}).
We note that this discussion can be extended to the cases of generalized $C_6$ symmetry.

From these results, we obtain the relation
\begin{equation}
 \label{eq:11}
  \begin{split}
   (-1)^{\nu_{C_4}}
   &=e^{i \pi \sum_{i=1}^4 p_i} \\
   &= e^{i \left[  -\arg \Delta (\Gamma)+ \arg \Delta (M)   \right]} \\
   &= \text{sgn} \Delta (\Gamma) \text{sgn} \Delta (M).
  \end{split}
\end{equation}
Here, we used the fact that $\Delta (\bm{k})$ is real at $\Gamma$ and $M$ points because of Eq.~(\ref{eq:45}).
Therefore, we obtain the discriminant indicator with generalized $C_4$ symmetry in Eq.~(\ref{eq:43}).
We note that even if the exceptional points exist on the red line in Fig.~\ref{fig:BZ_GRS}(a),
we can avoid by the continuous deformation of the integration path~\cite{Kim2015}.
We also note that if $\Delta = 0$, exceptional points emerge at these points.

\subsection{Two-dimensional system with generalized sixfold rotational symmetry\label{sec:two-dimens-syst-1}}
In order to derive Eq.~(\ref{eq:44}), let us consider the discriminant number composed along the closed path illustrated in Fig.~\ref{fig:BZ_GRS}(b).
We decompose the path for the integral into four paths $p_i$ ($i=1, \cdots, 4$)
in a similar way to the previous section [see Eq.~(\ref{eq:13})].

Firstly, we discuss the integration of $p_2 + p_3$ in a similar way to the previous case [see Eq.~(\ref{eq:14})].
In the presence of the generalized $C_6$ symmetry,
the integration of $\Delta (\bm{k})$ along $\Gamma_2$
is mapped to the one of $\Delta^{\ast}(\bm{k})$ along $\Gamma_2^{\prime}$.
Additionally, from the periodicity in the BZ, this integration is mapped to one along the opposite direction of $\Gamma_3$.
Thus, in a similar way to the previous case, $p_{2}+p_3$ is simplified as
\begin{equation}
 \label{eq:16}
 \begin{split}
  & 2\pi i(p_{2} +p_{3}) \\
  & = \int_{\Gamma_2}  d\bm{k} \cdot \bm{\nabla}_{\bm{k}} \ln \Delta^{\ast} (R_6^{3} \bm{k})
   + \int_{\Gamma_3}    d\bm{k} \cdot \bm{\nabla}_{\bm{k}} \ln \Delta ( \bm{k}) \\
   & =\int_{\Gamma_2^{\prime}}   d\bm{k} \cdot \bm{\nabla}_{\bm{k}} \ln \Delta^{\ast} ( \bm{k})
   + \int_{\Gamma_3}   d\bm{k} \cdot \bm{\nabla}_{\bm{k}} \ln \Delta (\bm{k}) \\
   & = -\int_{\Gamma_3}    d\bm{k} \cdot \bm{\nabla}_{\bm{k}} \ln \Delta^{\ast} ( \bm{k})
   + \int_{\Gamma_3}  d\bm{k} \cdot \bm{\nabla}_{\bm{k}} \ln \Delta (\bm{k}) \\
  & = 2i \left[ \arg \Delta (K^{\prime})- \arg \Delta (M)+2\pi N_0 \right] ,
 \end{split}
\end{equation}
with some integer $N_0$.

Secondly, let us simplify $p_1 + p_4$.
The integration of $\Delta (\bm{k})$ along $\Gamma_1$
is mapped to the integration of $\Delta^{\ast}(\bm{k})$ along the opposite direction of $\Gamma_4$
by the generalized $C_6$ operator.
Hence, the integration $p_1 + p_4$ is
\begin{equation}
 \label{eq:17}
  2\pi i( p_{1} +p_{4})
   =  2i \left[ \arg \Delta (\Gamma)- \arg \Delta (K^{\prime})+2\pi N_0^{\prime}  \right],
\end{equation}
with some integer $N_0^{\prime}$.

Putting Eqs.~(\ref{eq:16}) and (\ref{eq:17}) together, we obtain Eq.~(\ref{eq:44}).
Here, we used the fact that $\Delta (\bm{k})$ is real at the $\Gamma$ and $M$ points because of Eq.~(\ref{eq:45}).

\subsection{Three-dimensional system
  \label{sec:three-dimens-syst}}
Let us see the discriminant number in the three-dimensional BZ and exceptional points from loops.
In the presence of generalized $C_{4}$ symmetry about the $z$axis,
the following indicator captures the exceptional loops crossing one of the planes at $k_z=0$ and $\pi$;
\begin{equation}
 \label{eq:19}
     \xi_{C_4,3D} = \text{sgn} \left[\Delta (\Gamma) \Delta (M) \Delta (Z) \Delta (A)\right].
\end{equation}
Here, $\Gamma$, $M$, $Z$, and $A$ denote high-symmetry points as shown in Fig.~\ref{fig:BZ_GRS}(c).
We note that $\Delta (\Gamma)$, $\Delta (M)$, $\Delta (Z)$, and $\Delta (A)$ are real because of Eq.~(\ref{eq:41}).
All possible configurations of $\text{sgn}\Delta (\bm{k})$ giving $\xi_{C_n,3D} = -1$ are shown in Appendix~\ref{sec:spec-conf-with}.
The above indicator is obtained by applying the argument in Sec.~\ref{sec:two-dimens-syst}
to the two-dimensional subsystems at $k_z=0$ and $\pi$
where generalized $C_4$ symmetry is preserved.
In a similar way, we can introduce the indicator for systems with generalized $C_6$ symmetry,
\begin{equation}
 \label{eq:20}
   \xi_{C_6,3D} = \text{sgn} \left[\Delta (\Gamma) \Delta (M) \Delta (A) \Delta (L)\right],
\end{equation}
where $\Gamma$, $M$, $A$, and $L$ are high-symmetry points in Fig.~\ref{fig:BZ_GRS}(d).
We note that $\Delta (\Gamma)$, $\Delta (M)$, $\Delta (A)$, and $\Delta (L)$ are real because of Eq.~(\ref{eq:41}).

\section{Applications to toy models\label{sec:toy-model-with}}
In this section, we demonstrate that the discriminant indicator
with generalized $C_n$ symmetry ($n=4, \ 6$) predicts the existence of the exceptional points and loops
by numerically analyzing the two- and three-dimensional toy models with generalized $C_n$ symmetry.

\subsection{Two-dimensional toy model with generalized fourfold rotational symmetry\label{sec:toy-model-with-1}}
\begin{figure}[t]
 \centering
\includegraphics[trim={0cm 0cm 0cm 0cm},width =\hsize]{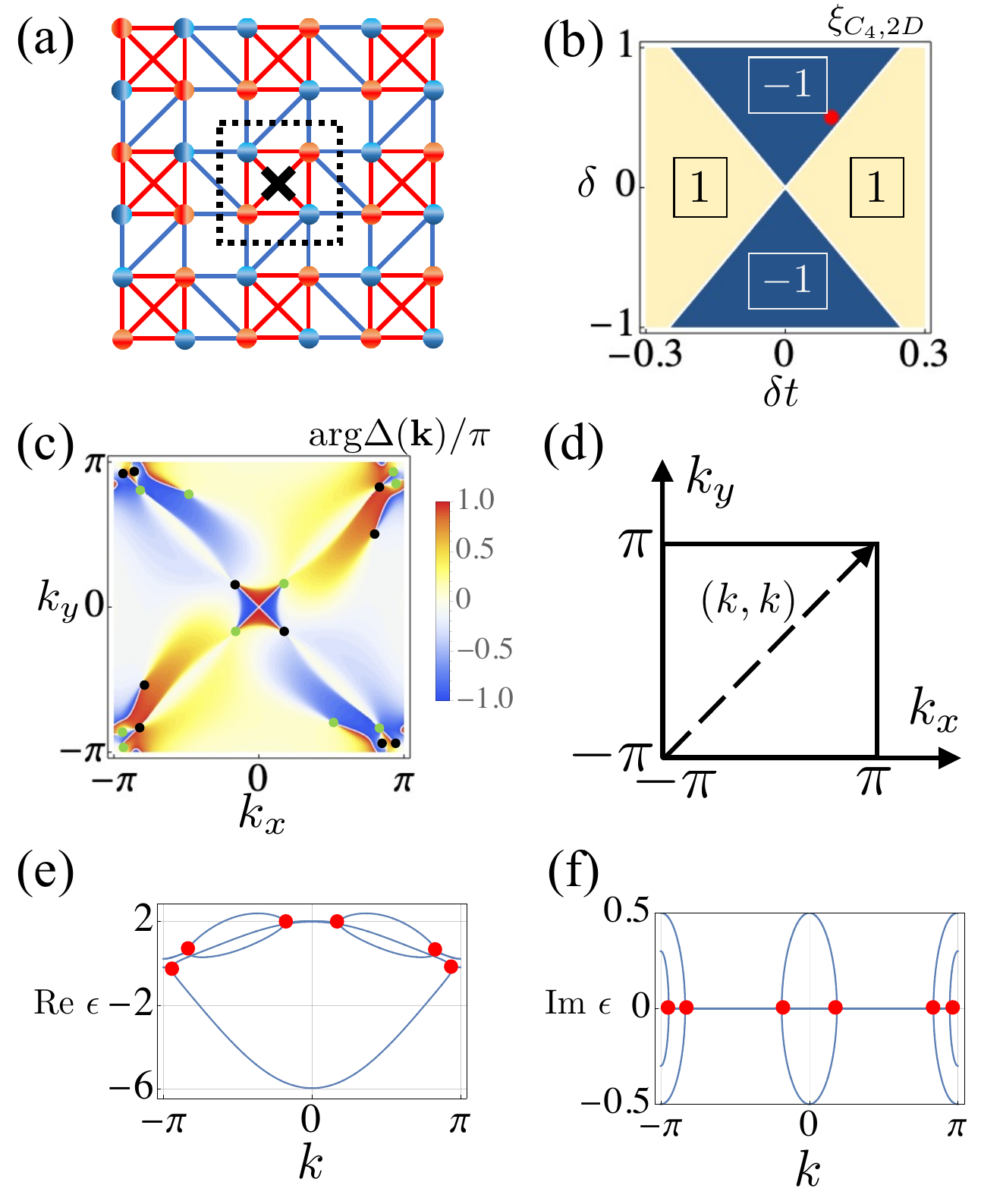}
\caption{(Color online)
 (a) Schematic figure of the toy model with generalized $C_4$ symmetry.
 The unit cell is enclosed by the black dashed box.
 The cross denotes the center of the generalized fourfold rotation.
 (b) Color plot of the map of the discriminant indicator $\xi_{C_4,2D}$.
 In the yellow regions (blue regions), the indicator takes a value of $1$ ($-1$).
 (c) Color map of $\arg \Delta (\bm{k})/\pi$ against $k_x$ and $k_y$ for $(\delta  ,\ \delta t)=(0.1,\ 0.5)$ [see the red dot in (b)].
 Green and black dots represent exceptional points with $\nu = 1$ and $-1$, respectively.
 (d) Sketch of the BZ.
 (e) and (f) The real and imaginary band structures  for $(\delta t , \ \delta  )= ( 0.1,\ 0.5)$, respectively.
 We plot along the line $k_x = k_y = k$ in (d).
 Red dots denote exceptional points in the band structure.
 }
 \label{fig:TOY_GFRS}
\end{figure}

Let us consider the two-dimensional toy model with generalized $C_4$ symmetry in Fig.~\ref{fig:TOY_GFRS}(a),
whose Hamiltonian is defined in Appendix~\ref{sec:deta-hamilt-toy}.
In Fig.~\ref{fig:TOY_GFRS}(a),
the red (blue) dots denote the on-site potentials $i \delta $ ($-i \delta$).
The red (blue) lines denote the hoppings $1 + \delta t $ ($1- \delta t$).

The above model respects generalized $C_4$ symmetry whose rotation axis is illustrated by a cross in Fig.~\ref{fig:TOY_GFRS}(a).
The explicit form of $U_{C_4}$ is written in Appendix~\ref{sec:deta-hamilt-toy}.

The numerical results for discriminant indicators are displayed in Fig.~\ref{fig:TOY_GFRS}(b).
The discriminant indicator takes vales of $\xi_{C_4,2D}=1$ and $-1$ in yellow and blue regions, respectively.
We note that $\xi_{C_4,2D}$ vanishes for $\delta t =0$ or $\delta = 0$
since $\Delta (\bm{k})$ vanishes at the high-symmetry point in the BZ.
Additionally, at the boundary of the phases, $\xi_{C_4,2D}$ vanishes for the same reason.
For $\xi_{C_4,2D}=-1$, the exceptional points can be observed by computing $\text{arg}\Delta (\bm{k}) / \pi$.
As an example, we show the map of $\text{arg}\Delta (\bm{k}) / \pi$ in momentum space
for $(\delta, \ \delta t) = (0.1, \ 0.5)$ [see Fig.~\ref{fig:TOY_GFRS}(c)].
Green and black dots in Fig.~\ref{fig:TOY_GFRS}(c) represent exceptional points characterized by $\nu = 1$ and $-1$, respectively.
This result indicates that the discriminant number along the path in Fig.~\ref{fig:BZ_GRS}(a) takes a value of $-1$.

Figures~\ref{fig:TOY_GFRS}(e) and \ref{fig:TOY_GFRS}(f) display the band structure along the line $k_x = k_y = k$ illustrated in Fig.~\ref{fig:TOY_GFRS}(d).
Figs.~\ref{fig:TOY_GFRS}(e) and \ref{fig:TOY_GFRS}(f), we can find exceptional points.

\subsection{Two-dimensional toy model with generalized sixfold rotational symmetry\label{sec:toy-model-with-2}}
\begin{figure}[t]
 \centering
\includegraphics[trim={0cm 0cm 0cm 0cm},width =\hsize]{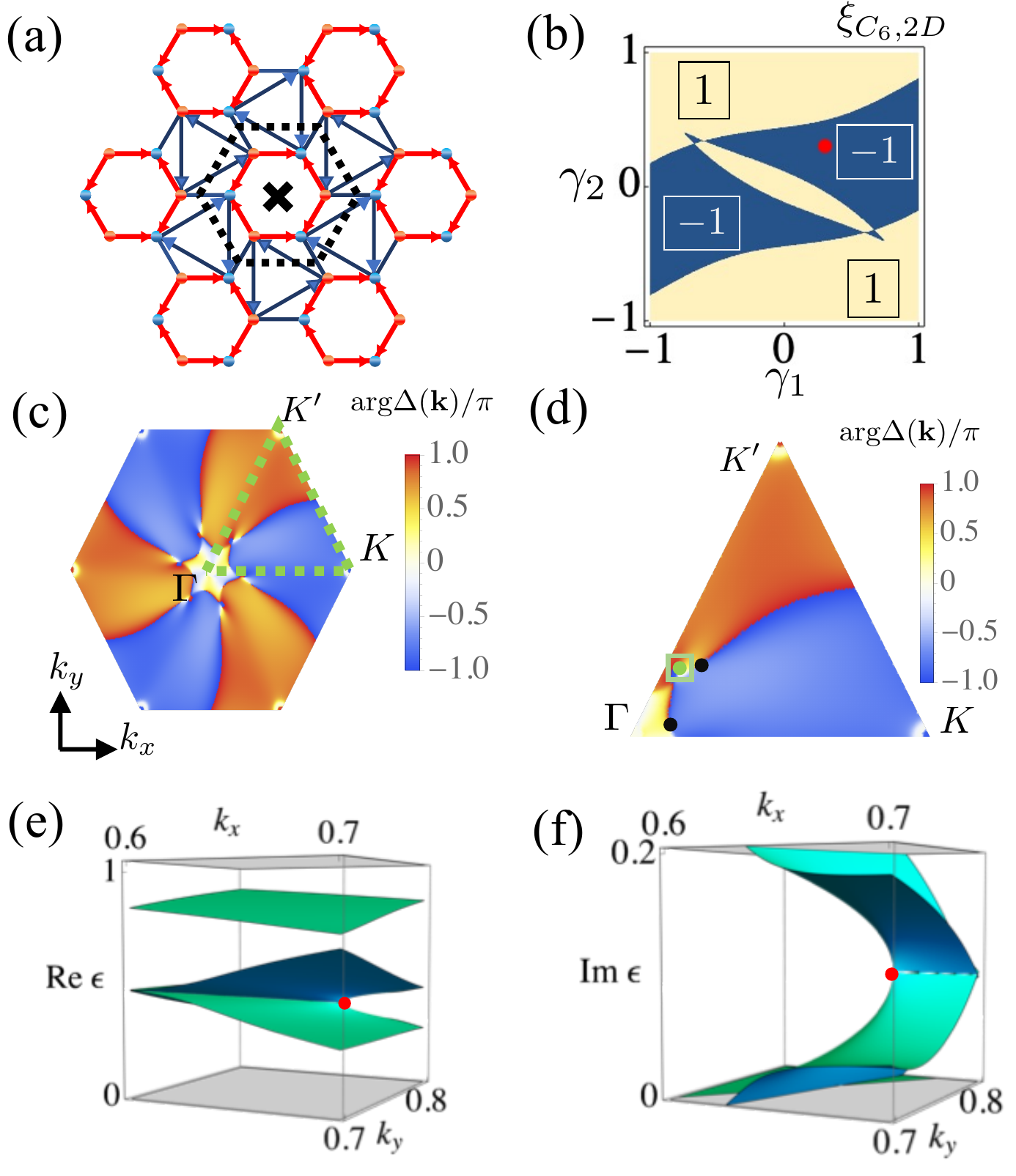}
\caption{(Color online)
 (a) Schematic figure of the toy model with generalized $C_6$ symmetry.
 The unit cell is enclosed by the black dashed hexagon.
 The cross denotes the center of the generalized sixfold rotation.
 (b) Color plot of the map of the discriminant indicator $\xi_{C_6,2D}$ for $\delta = 0.1$.
 In the yellow regions (blue regions), the indicator takes a value of $1$ $(-1)$.
 (c) Color map of $\arg \Delta (\bm{k})/\pi$ for $(\delta  ,\ \gamma_1 , \ \gamma_2)=(0.1,\ 0.3, \ 0.3)$ [see the red dot in panel (b)] in the BZ.
 (d) Magnified version of panel (c) in the region enclosed by a green dashed triangle.
 (e) and (f) The real and imaginary band structures for $0.6 \leq k_x \leq 0.7$ and $0.7 \leq k_y \leq 0.8$
 [see the green box in (d)].
 The red point in the band structure represents an exceptional point.
 }
 \label{fig:TOY_GSRS}
\end{figure}

We consider the two-dimensional toy model with generalized $C_6$ symmetry in Fig.~\ref{fig:TOY_GSRS}(a),
whose Hamiltonian is defined in Appendix~\ref{sec:deta-hamilt-toy}.
In this figure, the red (blue) dots denote the on-site potentials $i \delta $ ($- i \delta$).
The red (blue) lines denote the non-reciprocal hoppings.
Specifically, the hopping along the red (blue) arrows is $1+\gamma_1$ ($1+\gamma_2$).
The hopping opposite to the red (blue) arrows is $1-\gamma_1$ ($1-\gamma_2$).
For the blue lines without arrows, the hopping is $1$ for both directions.

The above model respects generalized $C_6$ symmetry whose rotation axis is illustrated by a cross in Fig.~\ref{fig:TOY_GSRS}(a).
The explicit form of $U_{C_6}$ is written in Appendix~\ref{sec:deta-hamilt-toy}.

We show the numerical result for the discriminant indicator $\xi_{C_6,2D}$ in Eq.~(\ref{eq:44}) for $\delta = 0.1$
[see Fig.~\ref{fig:TOY_GSRS}(b)].
In Fig.~\ref{fig:TOY_GSRS}(b), there exist two phases where the discriminant indicator takes values of $\xi_{C_6,2D}=1$ (yellow area) and $-1$ (blue area).
We note that $\xi_{C_6,2D}$ vanishes for $\gamma_1 = 0$ and $\gamma_2 = 0$
since $\Delta (\bm{k})$ vanishes at the high-symmetry point in the BZ.
Additionally, at the boundary between phases, $\xi_{C_6,2D}$ vanishes for the same reason.
In the blue-colored region, the indicator predicts the presence of exceptional points.
To be concrete, we compute $\text{arg}\Delta (\bm{k}) / \pi$ in the momentum space
for $(\delta,\ \gamma_1,\ \gamma_2) = (0.1,\ 0.3, \ 0.3)$ [see Figs.~\ref{fig:TOY_GSRS}(c) and \ref{fig:TOY_GSRS}(d)].
Green and black dots in Fig.~\ref{fig:TOY_GSRS}(d) represent exceptional points characterized by $\nu = 1$ and $-1$, respectively.
Figure~\ref{fig:TOY_GSRS}(d) indicates that the discriminant number computed along the path in Fig.~\ref{fig:BZ_GRS}(b) takes a value of $-1$.

Figures~\ref{fig:TOY_GSRS}(e) and \ref{fig:TOY_GSRS}(f) plot the band structure for $(\delta , \gamma _1 , \gamma_2) = (0.1, \ 0.3,\ 0.3)$,
which indicates the emergence of the exceptional point at the point denoted by the red dots in Figs.~\ref{fig:TOY_GSRS}(e) and~\ref{fig:TOY_GSRS}(f).

\subsection{Three-dimensional toy model with generalized fourfold rotational symmetry\label{sec:three-dimens-toy-1}}
\begin{figure}[t]
 \centering
\includegraphics[trim={0cm 0cm 0cm 0cm},width =\hsize]{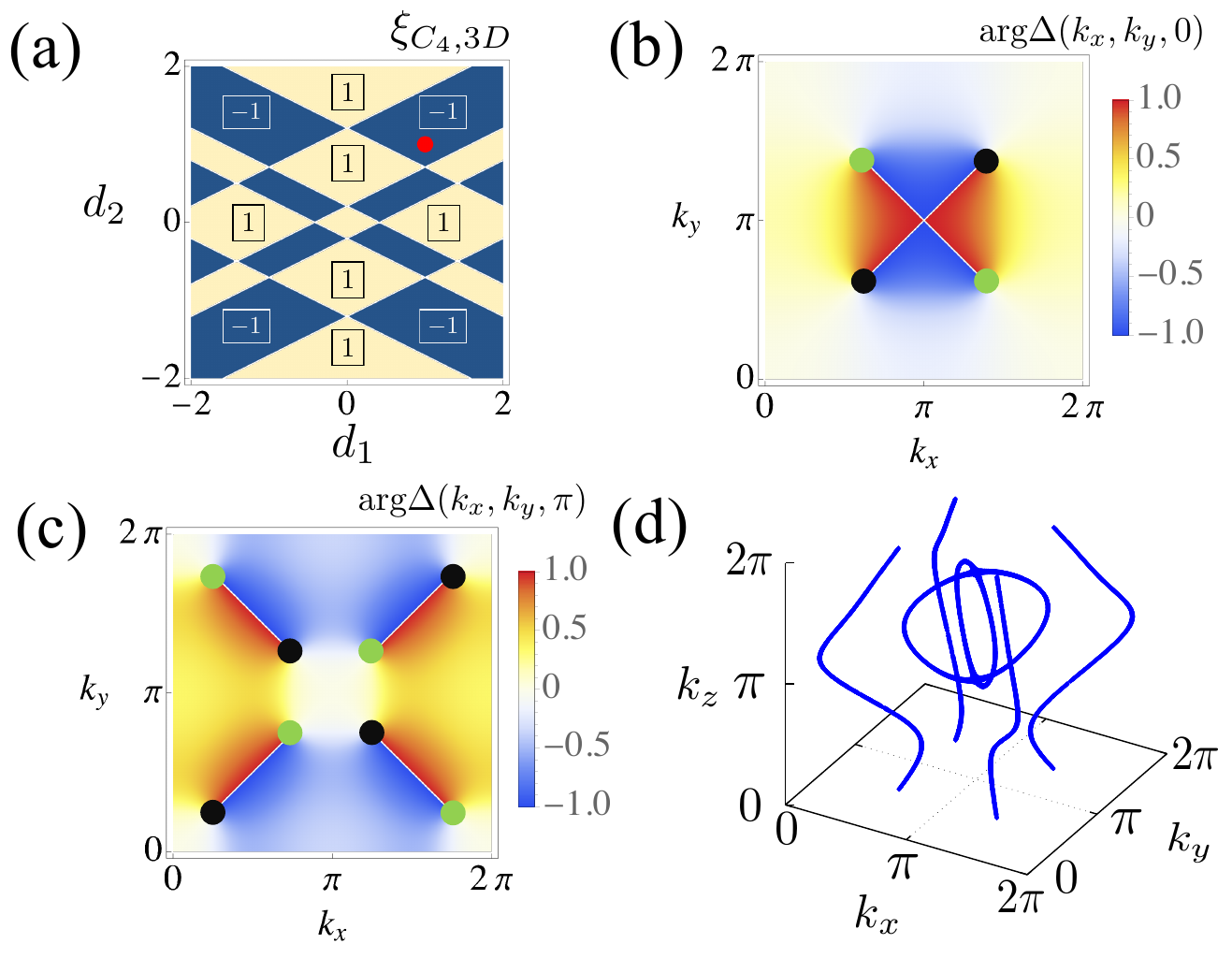}
\caption{(Color online)
 (a) Color plot of the map of the discriminant indicator $\xi_{C_4,3D}$.
 In the yellow regions (blue-colored regions), the indicator takes a value of $1$ $(-1)$.
 (b)-(c) Color maps of $\text{arg}\Delta(\bm{k})$ for $(d_1,d_2)= (1,1)$ in the BZ [see the red dot in (a)].
 (b) and (c) are displayed at the plane for $k_z=0$ and $k_z = \pi$ in the BZ, respectively.
 Green and black dots represent the exceptional loops crossing the plane for fixed $k_z$ with $\nu=1$ and $\nu=-1$, respectively.
 (d) Momentum points $\bm{k}$ satisfying $\text{abs}\Delta (\bm{k}) < 0.05$.
 }
\label{fig:TOY_GFRS_3d}
\end{figure}

Consider the three-dimensional toy model with generalized fourfold rotational symmetry whose
Hamiltonian reads
\begin{equation}
 \label{eq:5}
  H(\bm{k}) =
  \sum_{i=1}^3 R_i (\bm{k}) \sigma_i,
\end{equation}
with
\begin{subequations} % 2022-05-14 11:21の式群
\begin{align}
   R_1 (\bm{k}) &= i \cos k_z,   \label{eq:12}\\
   R_2 (\bm{k}) &= (\cos k_x - \cos k_y) + i,   \label{eq:32}\\
   R_3(\bm{k})  &= d_1 \cos k_z + \sin k_z + d_2 (\cos k_x + \cos k_y)+1.  \label{eq:33}
\end{align}
\end{subequations}
Here, $\sigma_i$ ($i=1$, $2$, $3$) denote the Pauli matrices.
The above model respects generalized $C_4$ symmetry about the $z$axis with $U_{C_4}= \sigma_3$.

Figure~\ref{fig:TOY_GFRS_3d}(a) displays the numerical result for the discriminant indicator $\xi_{C_4,3D}$.
In Fig.~\ref{fig:TOY_GFRS_3d}(a), there exist two phases where the discriminant indicator
takes values of $\xi_{C_4,3D} = 1$ (yellow area) and $\xi_{C_4,3D} = -1$ (blue area).
For $\xi_{C_4,3D}=-1$, the exceptional loops crossing the plane for $k_z = 0$ or $\pi$ can be obtained by computing the map of $\text{arg}\Delta (\bm{k})/\pi$.
As an example, we show the maps of $\text{arg}\Delta (k_x,k_y,0)$ and $\text{arg}\Delta (k_x,k_y,\pi)$ in momentum space for $(d_1,d_2)=(1,1)$ [see Figs.~\ref{fig:TOY_GFRS_3d}(b) and~\ref{fig:TOY_GFRS_3d}(c)].
In Figs.~\ref{fig:TOY_GFRS_3d}(b) and \ref{fig:TOY_GFRS_3d}(c), green and black dots represent exceptional loops characterized by $\nu=1$ and $-1$, respectively.
By comparing these figures, we obtain the emergence of the exceptional loops crossing the plane for $k_z = \pi$.

Figure~\ref{fig:TOY_GFRS_3d}(d) displays the momentum points $\bm{k}$ satisfying $\text{abs}\Delta (\bm{k}) < 0.05$,
indicating the emergence of exceptional loops.
We note that exceptional loops predicted by the nontrivial value of $\xi_{C_4,3D}$ merge at the line specified by $(k_x,k_y)=(0,0)$ or $(\pi, \pi)$,
originating from the symmetry constraint on the exceptional points for fixed $k_z$ in three dimensions
(for more details see Appendix~\ref{sec:merg-except-loops}).

\begin{figure}[t]
 \centering
\includegraphics[trim={0cm 0cm 0cm 0cm},width =\hsize]{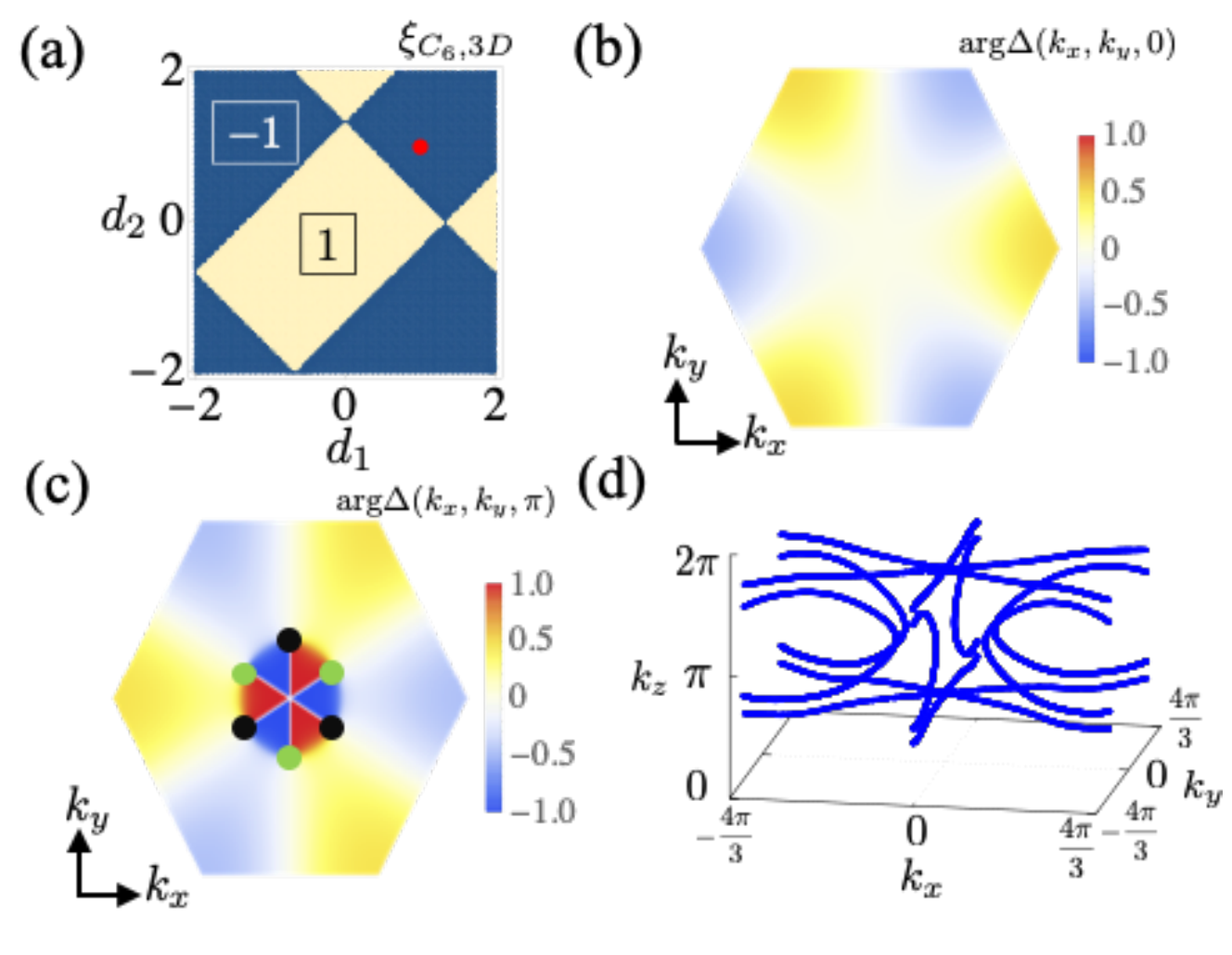}
\caption{(Color online)
 (a) Color plot of the map of the discriminant indicator $\xi_{C_6,3D}$.
 In the yellow regions (blue regions), the indicator takes a value of $1$ $(-1)$.
 (b) and (c) Color maps of $\text{arg}\Delta(\bm{k})$ for $(d_1,d_2)= (1,1)$ [see the red dot in panel (a)] in the BZ.
 Panel (b) and Panel (c) are displayed at the plane for $k_z=0$ and $k_z = \pi$ in the BZ, respectively.
 Green and black dots represent the exceptional loops crossing the plane for fixed $k_z$ with $\nu=1$ and $\nu=-1$, respectively.
 (d) Momentum points $\bm{k}$ satisfying $\text{abs}\Delta (\bm{k}) < 0.05$.
 }
\label{fig:TOY_GSRS_3d}
\end{figure}

\subsection{Three-dimensional toy model with generalized sixfold rotational symmetry\label{sec:three-dimens-toy-2}}

Consider the three-dimensional toy model with generalized sixfold rotational symmetry whose
Hamiltonian reads
\begin{equation}
\label{eq:35}
  H(\bm{k}) =
  \sum_{i=1}^3 R_i (\bm{k}) \sigma_i,
\end{equation}
with
\begin{subequations}
 \begin{align}
  R_1 (\bm{k}) &= id_1 C(k_x,k_y) + i \cos k_z+i,  \label{eq:37}\\
  R_2 (\bm{k}) &= S(k_x,k_y) + i \sin k_z, \label{eq:46}\\
  R_3(\bm{k}) &=  d_2 C(k_x,k_y) +2 i S(k_x,k_y) \nonumber \\
  &\quad + 5 \cos k_z + \sin k_z + 1   \label{eq:47} .
 \end{align}
\end{subequations}
Here, $C(k_x,k_y)$ and $S(k_x,k_y)$ are defined as
\begin{subequations}
 \begin{align}
  C (k_x , k_y) &= \cos (-k_x)+ \cos \left( k_x/2 +\sqrt{3} k_y/2  \right) \nonumber \\
  &\quad + \cos \left( k_x/2 - \sqrt{3} k_y /2  \right)  ,   \label{eq:48} \\
  S (k_x , k_y) &= \sin (-k_x)+ \sin \left( k_x /2 + \sqrt{3} k_y/2  \right)\nonumber \\
  & \quad + \sin \left( k_x/2 - \sqrt{3} k_y/2  \right) .\label{eq:15}
 \end{align}
\end{subequations}
The above model respects generalized $C_6$ symmetry about the $z$axis with $U_{C_6}= \sigma_3$.

Figure~\ref{fig:TOY_GSRS_3d}(a) displays the numerical result of the discriminant $\xi_{C_6,3D}$.
In Fig.~\ref{fig:TOY_GSRS_3d}(a), there exist two phases where the discriminant indicator
takes values of $\xi_{C_6,3D} = 1$ (yellow area) and $\xi_{C_6,3D} = -1$ (blue area).
For $\xi_{C_6,3D}=-1$, the exceptional loops crossing the plane for $k_z = 0$ or $\pi$ can be obtained by computing the map of $\text{arg}\Delta (\bm{k})/\pi$.
As an example, we show the maps of $\text{arg}\Delta (k_x,k_y,0)$ and  $\text{arg}\Delta (k_x,k_y,\pi)$ in the momentum space for $(d_1,d_2)=(1,1)$ [see Figs.~\ref{fig:TOY_GSRS_3d}(b) and~\ref{fig:TOY_GSRS_3d}(c)].
In Fig.~\ref{fig:TOY_GSRS_3d}(c), green and black dots represent exceptional loops characterized by $\nu=1$ and $-1$, respectively.
By comparing Fig.~\ref{fig:TOY_GSRS_3d}(b) with \ref{fig:TOY_GSRS_3d}(c), we obtain the emergence of the exceptional loops crossing the plane for $k_z = \pi$.

Figure~\ref{fig:TOY_GSRS_3d}(d) displays the momentum points $\bm{k}$ satisfying $\text{abs}\Delta (\bm{k}) < 0.05$,
indicating the emergence of exceptional loops.
We note that, in contrast to the case of generalized $C_4$ symmetry in Sec.~\ref{sec:three-dimens-toy-1}, generalized $C_6$ symmetry does not predict the merging of the exceptional loops
since the exceptional loops can cross the boundary of the BZ
(for more details see Appendix~\ref{sec:merg-except-loops}).

\section{Topoelectrical systems with generalized fourfold rotational symmetry\label{sec:topo-syst-with}}
\begin{figure}[t]
 \centering
\includegraphics[trim={0cm 0cm 0cm 0cm},width =0.6\hsize]{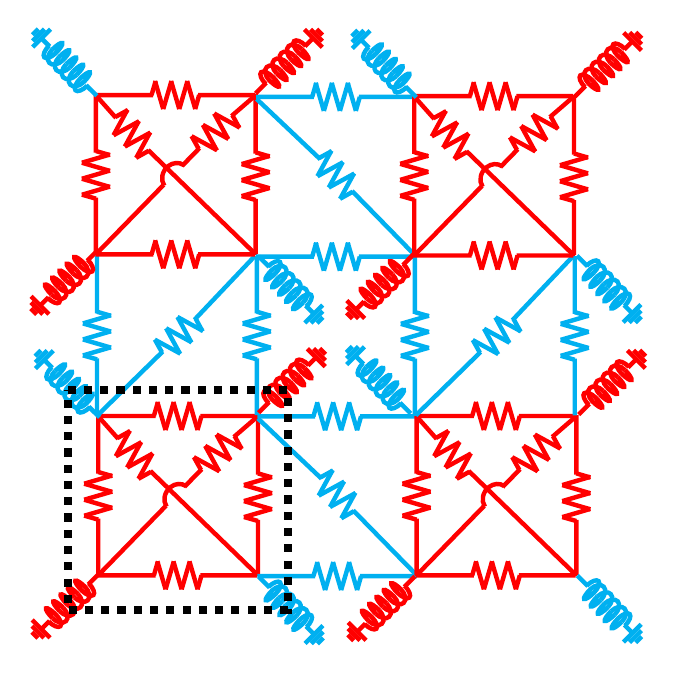}
\caption{(Color online)
 Schematic figure of the topoelectrical system with generalized $C_4$ symmetry.
 The unit cell is enclosed by the dashed black box.
 We omit power sources connecting each nodes.
 }
 \label{fig:TEC_GFRS}
\end{figure}

We have shown that our indicators capture the exceptional points and loops in toy models.
In this section, we discuss the relevance of generalized $C_4$ symmetry to the topoelectrical circuits~\cite{Albert2015,Lee2018,Hofmann2020}.

Consider the two-dimensional topoelectrical circuit with generalized $C_4$ symmetry in Fig.~\ref{fig:TEC_GFRS}.
To describe this system, under the periodic boundary condition,
we define the voltages at each node as $\bm{V}(\bm{k}, \omega)$
and electric currents between the node and the power source as $\bm{I}(\bm{k}, \omega)$.
Here, $\omega$ denotes the frequency of the currents from the power source into the node.
The admittance of the red (blue) resistors shown in Fig.~\ref{fig:TEC_GFRS} is $R_{\mathrm{r}}$ ($R_{\mathrm{b}}$)
with $R_{\mathrm{r}} = 1+\delta r$ and $R_{\mathrm{b}} = 1- \delta r$.
The admittance of red (blue) inductors is
$-i \omega^{-1} L_{\mathrm{r}}^{-1}$ ($-i \omega^{-1} L_{\mathrm{b}}^{-1}$), with $L_{\mathrm{r}}=(1+ \delta )^{-1}$ and $L_{\mathrm{b}}=(1- \delta)^{-1} $.
The relation between $\bm{I} (\bm{k}, \omega)$ and $\bm{V}(\bm{k}, \omega)$ is
\begin{equation}
 \label{eq:9}
  \bm{I}  (\bm{k},\omega) = J (\bm{k},\omega) \bm{V}(\bm{k},\omega).
\end{equation}
Here, $J(\bm{k},\omega)$ denotes the admittance matrix,
\begin{equation}
 \label{eq:26}
  \begin{split}
   J(\bm{k},\omega) &= 6\1_{4 \times 4}
   + (i \omega)^{-1} \text{diag} ( L_{\mathrm{b}}^{-1},  L_{\mathrm{b}}^{-1},  L_{\mathrm{r}}^{-1},  L_{\mathrm{r}}^{-1} ) \\
   &-
  \begin{pmatrix}
    0                & R_{\mathrm{r}}   & R_{\mathrm{r}} &R_{\mathrm{r}} \\
   R_{\mathrm{r}}    & 0                                 & R_{\mathrm{r}}      & R_{\mathrm{r}}  \\
   R_{\mathrm{r}}    & R_{\mathrm{r}}   & 0             & R_{\mathrm{r}}  \\
   R_{\mathrm{r}}    & R_{\mathrm{r}}   & R_{\mathrm{r}}  &  0
  \end{pmatrix} \\
      &-
  \begin{pmatrix}
    0                          & R_{\mathrm{b}}e^{i k_x}    & R_{\mathrm{b}}e^{ik_x}      & R_{\mathrm{b}}e^{i k_y} \\
   R_{\mathrm{b}}e ^{- i k_x}  & 0                          & R_{\mathrm{b}}e^{-ik_y}     &  R_{\mathrm{b}}e^{-i k_x} \\
   R_{\mathrm{b}}e^{-i k_x}    & R_{\mathrm{b}}e^{ i k_y}   & 0                           & R_{\mathrm{b}}e^{ik_y} \\
   R_{\mathrm{b}}e^{-ik_y}     & R_{\mathrm{b}}e^{i k_x}    & R_{\mathrm{b}}e ^{ -i k_y}  &  0
  \end{pmatrix}.
  \end{split}
\end{equation}
We note that the constant diagonal terms just shift the band structure and do not contribute to the emergence of exceptional points.
Thus, we define $J^{\prime}(\bm{k},\omega)$ by removing the constant diagonal terms of the admittance matrix,
\begin{equation}
 \label{eq:28}
  J^{\prime} (\bm{k},\omega) =J (\bm{k},\omega)- [\text{Tr} J (\bm{k},\omega)] \1_{4 \times 4}/4.
\end{equation}
Clearly, $J^{\prime} (\bm{k},\omega)$ corresponds to the Hamiltonian in Sec.~\ref{sec:toy-model-with-1}.

\section{Summary\label{sec:summary}}
Extending the argument of the work in \cite{Yoshida2022},
we have introduced the discriminant indicators for two- and three-dimensional systems with generalized $C_n$ symmetry ($n=4$, $6$).
In two dimensions, the indicators can be computed from the parity of the discriminant at the $\Gamma$ and $M$ points [see Eqs.~(\ref{eq:43}) and (\ref{eq:44})],
which is in contrast to the case of generalized inversion symmetry.
These indicators taking a nontrivial value predict the emergence of exceptional points and exceptional loops without the input of the reference energy.
We have numerically confirmed that these indicators capture the exceptional points and the loops for two- and three-dimensional toy models.

\acknowledgements
The authors thank R. Okugawa for fruitful discussion.
This work is supported by JSPS KAKENHI Grants No. JP17H06138, No. JP20H04627, and No. JP21K13850
and also by JST CREST, Grant No. JPMJCR19T1, Japan.

\appendix

\begin{figure}[t]
 \centering
\includegraphics[trim={0cm 0cm 0cm 0cm},width =0.75\hsize]{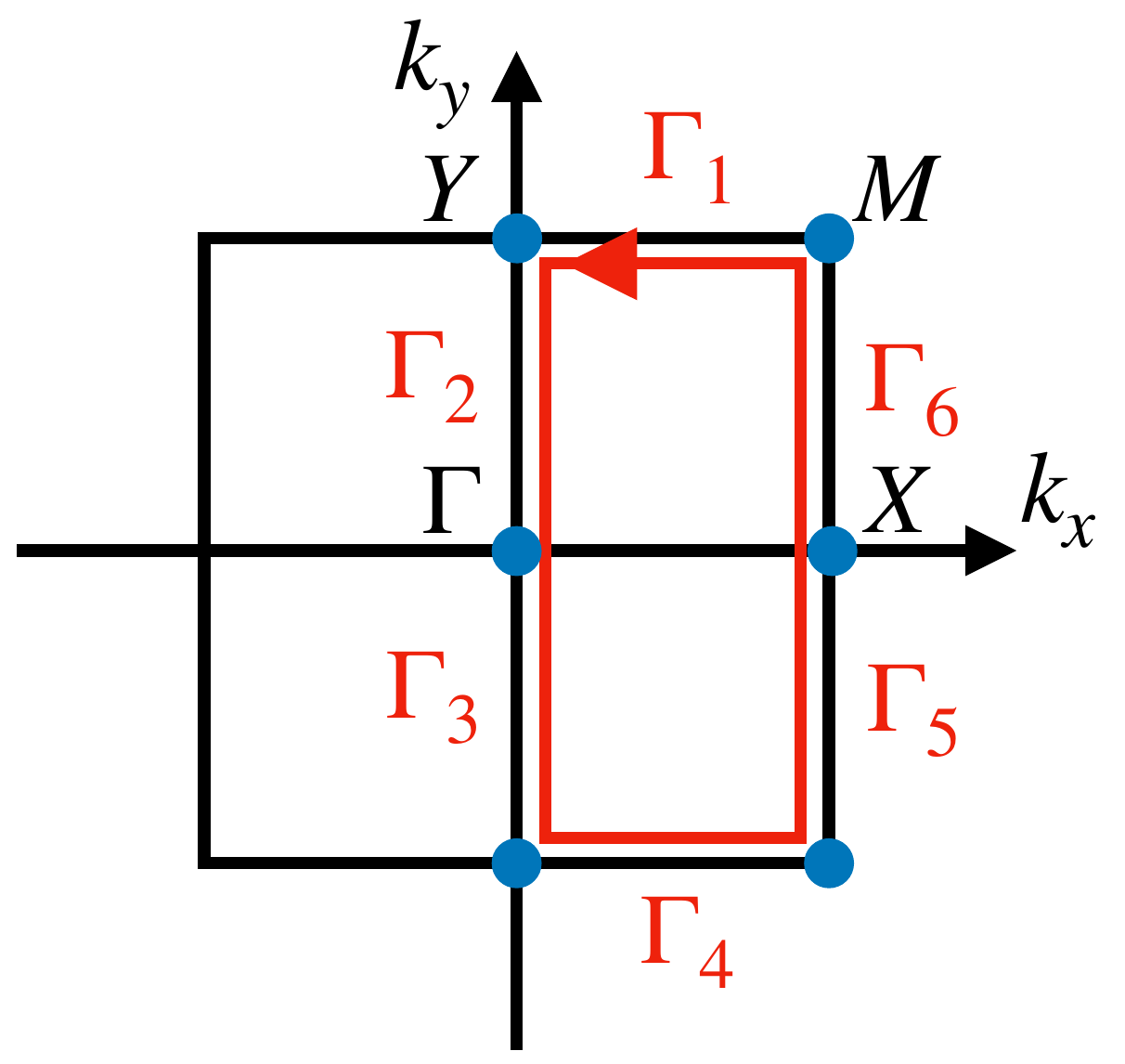}
\caption{(Color online)
 Sketch of the BZ with generalized $C_2$ symmetry and the path of integrations in Eq.~(\ref{eq:13}).
 The high-symmetry points are represented by blue dots.
 The path $\Gamma_i$ ($i=1,\cdots,6$) connects the high-symmetry points.
 }
\label{fig:BZ_GTRS}
\end{figure}

\section{Discriminant indicator with generalized twofold rotational symmetry\label{sec:discr-numb-with}}

We briefly discuss the indicators for systems with generalized $C_2$ symmetry, although the problem is reduced to the one for generalized inversion symmetry.

With generalized $C_2$ symmetry in two-dimensional systems,
the discriminant number is simplified to the computation at the high-symmetry points in the BZ~\cite{Yoshida2022}.
This symmetry enables us to rewrite Eq.~(\ref{eq:2}) as the discriminant indicator,
\begin{equation}
 \label{eq:4}
  (-1)^{\nu_{C_2}}:=\xi_{C_2,2D} =  \text{sgn}\left[\Delta (\Gamma)\Delta (X)\Delta (M)\Delta (Y) \right],
\end{equation}
where
$\nu_{C_2}$ denotes the discriminant number defined along the path in Fig.~\ref{fig:BZ_GTRS}.
Symbols $\Gamma$, $X$, $M$, and $Y$ denote the high-symmetry points [see Fig.~\ref{fig:BZ_GTRS}].
We note that $\Delta (\Gamma)$, $\Delta (X)$, $\Delta (M)$, and $\Delta (Y)$ are real because of Eq.~(\ref{eq:45}).

We see the derivation of Eq.~(\ref{eq:4}).
Let us consider the discriminant number [see Eq.~(\ref{eq:1})] composed along the closed path illustrated in Fig.~\ref{fig:BZ_GTRS}.
In a way similar to what we did in Sec.~\ref{sec:two-dimens-syst}, we decompose the closed path into six parts.
The integral for the paths $\Gamma_i$ ($i=1,\cdots,6$) is denoted by $p_i$ ($i=1,\cdots,6$) in Eq.~(\ref{eq:13}).

We note that $p_1 = -p_4$ holds due to the periodicity of the BZ.
We also note that $p_2 + p_3$ is simplified as
\begin{equation}
 \label{eq:18}
 2\pi i(p_2 + p_3) = 2i \left[ \text{arg}\Delta (\Gamma)- \text{arg} \Delta (Y)+2\pi N_0   \right],
\end{equation}
with some integer $N_0$.
This is because $\Gamma_3$ is mapped to $\Gamma_2$ by applying the operator of generalized $C_2$ symmetry.
In a similar way, we have
\begin{equation}
\label{eq:3}
  2\pi i (p_5 + p_6) = 2i \left[ \text{arg}\Delta (M)- \text{arg} \Delta (X) +2\pi N_0^{\prime}  \right],
\end{equation}
with some integer $N_0^{\prime}$.

Putting Eqs.~(\ref{eq:18}) and (\ref{eq:3}) together, we obtain Eq.~(\ref{eq:4}).
Here, we have used the fact that $\Delta (\bm{k})$ is real at the $\Gamma$, $X$, $M$, and $Y$ points because of Eq.~(\ref{eq:45}).

\begin{figure}[b]
\label{fig:BZ_3d}
 \centering
\includegraphics[trim={0cm 0cm 0cm 0cm},width =\hsize]{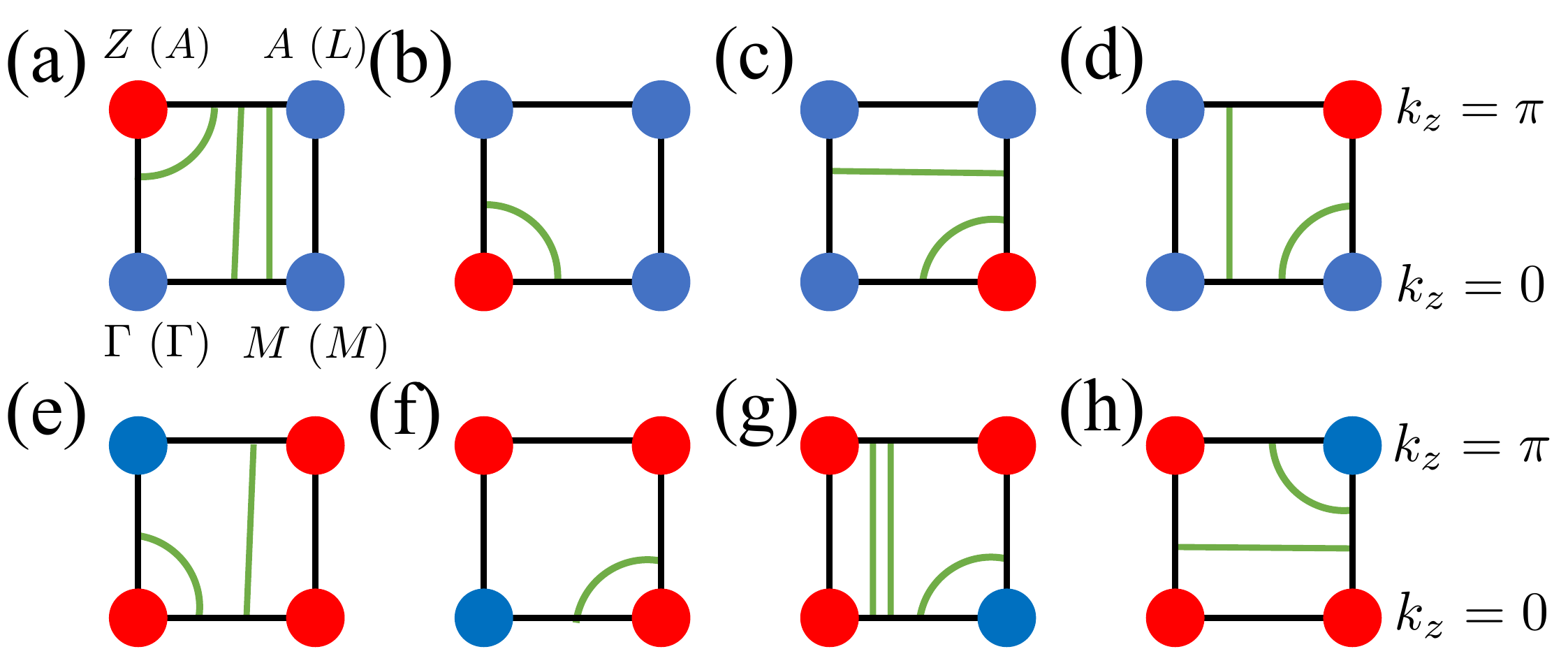}
\caption{(Color online)
 (a)$-$(h) The sketch of the configuration of the sign of the discriminant giving $\xi_{C_n,3D}=-1$.
 The squares whose vertices are $\Gamma$, $M$, $Z$, and $A$ ($\Gamma$, $M$, $L$, and $A$) denote
 the area in the BZ with generalized $C_4$ ($C_6$) symmetry.
 The blue (red) dots are the sign of the discriminant $\text{sgn} \Delta (\bm{k}) = 1$ ($-1$).
 Green lines represent the examples of exceptional loops.
 }
\end{figure}

\section{Specific configuration with the nontrivial value of the discriminant indicator\label{sec:spec-conf-with}}
In this appendix, we illustrate the specific configuration yielding $\xi_{C_n,3D}=-1$ (see Fig.~\ref{fig:BZ_3d}).
With this nontrivial value of the indicators, the exceptional loops emerge crossing one of the planes for $k_z=0$ or $\pi$ exist.
For example, in Fig.~\ref{fig:BZ_3d}(a), there exist an exceptional loop crossing the plane for $k_z = \pi$,
and two loops crossing both of the planes for $k_z=0$ and $\pi$.
In this case, at the plane for $k_z=0\ (\pi)$, the indicator for two-dimensional systems takes a trivial (nontrivial) value.
This result indicates that the discriminant indicator takes a value of $-1$.

\section{Details of the Hamiltonian in toy models\label{sec:deta-hamilt-toy}}
Here, we show the details of the Hamiltonian in Sec.~\ref{sec:toy-model-with}.
The bulk Hamiltonian describing the toy model in Fig.~\ref{fig:TOY_GFRS}(a) is
\begin{widetext}
 \begin{equation}
 \label{eq:30}
   H_{C_4} (\bm{k}) =
  \begin{pmatrix}
    i\delta                                    &  -(1+\delta t) - (1-\delta t) e^{-i k_x}   &  -(1+\delta t) - (1-\delta t) e^{-i k_x} &  -(1+\delta t) - (1-\delta t) e^{-i k_y} \\
    -(1+\delta t) - (1-\delta t) e^{i k_x}    &  i\delta                                   &  -(1+\delta t) - (1-\delta t) e^{i k_y}  &  -(1+\delta t) - (1-\delta t) e^{i k_x}  \\
    -(1+\delta t) - (1-\delta t) e^{i k_x}    &  -(1+\delta t) - (1-\delta t) e^{-i k_y}   & -i\delta                                 &  -(1+\delta t) - (1-\delta t) e^{-i k_y} \\
    -(1+\delta t) - (1-\delta t) e^{i k_y}    &  -(1+\delta t) - (1-\delta t) e^{-i k_x}   &  -(1+\delta t) - (1-\delta t) e^{i k_y}  & -i\delta
  \end{pmatrix}.
 \end{equation}

This model preserves generalized $C_4$ symmetry [see Eq.~(\ref{eq:6})] where $U_{C_4}$ is defined as
\begin{equation}
 \label{eq:31}
  U_{C_4}=
  \begin{pmatrix}
   0 & 0 & 0 & 1 \\
   0 & 0 & 1 & 0 \\
   1 & 0 & 0 & 0 \\
   0 & 1 & 0 & 0
  \end{pmatrix}.
\end{equation}

Additionally, the bulk Hamiltonian describing the toy model in Fig.~\ref{fig:TOY_GSRS}(a) is
\begingroup
\scriptsize
 \begin{equation}
\label{eq:34}
 \begin{split}
  &  H_{C_6} (\bm{k})  \\
  &=  \begin{pmatrix}
      i\delta      &  -(1 -\gamma_1)    &  -(1-\gamma_2) e^{i \bm{k} \cdot \bm{a}_2 } &  -  e^{i \bm{k} \cdot \bm{a}_2 }  & -(1+\gamma_2) e^{i \bm{k} \cdot (-\bm{a}_1 + \bm{a}_2)} & -(1-\gamma_1) \\
      -(1+\gamma_1) & - i \delta & -(1+\gamma_1) & -(1+\gamma_2) e^{i\bm{k} \cdot  (-\bm{a}_1 + \bm{a}_2)} &  -e^{i\bm{k} \cdot  (-\bm{a}_1 + \bm{a}_2)} & -(1-\gamma_2) e^{- i \bm{k} \cdot \bm{a}_1} \\
      -(1+\gamma_2)  e^{-i \bm{k} \cdot \bm{a}_2 } & -(1-\gamma_1) & i \delta & -(1-\gamma_1) & -(1-\gamma_2) e^{-i \bm{k} \cdot \bm{a}_1} & - e^{-i\bm{k} \cdot \bm{a}_1} \\
      - e^{-i \bm{k} \cdot \bm{a}_2} & -(1 - \gamma_2) e^{-i \bm{k} \cdot (-\bm{a}_1 + \bm{a}_2)} & -(1+\gamma_1) & - i\delta & - (1+\gamma_1) & -(1+\gamma_2) e^{-i \bm{k} \cdot \bm{a}_2} \\
      -(1-\gamma_2) e^{-i \bm{k} \cdot (-\bm{a}_1 + \bm{a}_2)} & -  e^{-i \bm{k} \cdot (-\bm{a}_1 + \bm{a}_2)} & -(1+\gamma_2) e^{i \bm{k} \cdot \bm{a}_1} & -(1-\gamma_1) & i \delta & - (1-\gamma_1) \\
      -(1+\gamma_1) & -(1+\gamma_2) e^{i \bm{k} \cdot \bm{a}_1} & -e^{i \bm{k} \cdot \bm{a}_1 } & -(1-\gamma_2) e^{i \bm{k} \cdot \bm{a}_2} & -(1+\gamma_1) & -i \delta
     \end{pmatrix}.
   \end{split}
 \end{equation}
\endgroup
\end{widetext}
Here, $\bm{a}_1=\left(\frac{1}{2},\frac{\sqrt{3}}{2}\right)$ and $\bm{a}_2=\left(-\frac{1}{2},\frac{\sqrt{3}}{2}\right)$ are the unit vectors.

This model preserves generalized $C_6$ symmetry [see Eq.~(\ref{eq:6})] where $U_{C_6}$ is defined as
\begin{equation}
 \label{eq:36}
  U_{C_6} =
  \begin{pmatrix}
   0 & 0 & 0 & 0 & 0 & 1 \\
   1 & 0 & 0 & 0 & 0 & 0 \\
   0 & 1 & 0 & 0 & 0 & 0 \\
   0 & 0 & 1 & 0 & 0 & 0 \\
   0 & 0 & 0 & 1 & 0 & 0 \\
   0 & 0 & 0 & 0 & 1 & 0
  \end{pmatrix}.
\end{equation}

\section{The merging of exceptional loops \label{sec:merg-except-loops}}
\begin{figure}[H]
 \centering
\includegraphics[trim={0cm 0cm 0cm 0cm},width =\hsize]{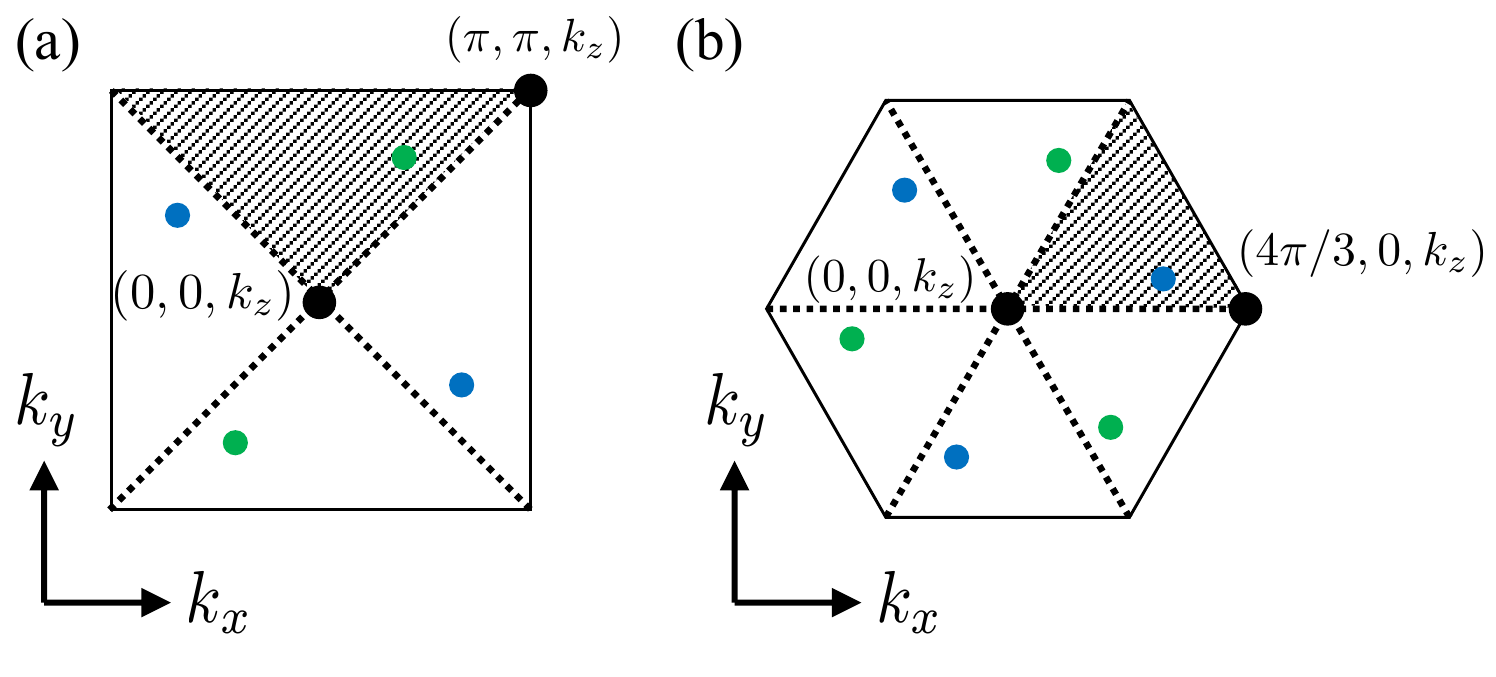}
\caption{(Color online)
 (a) and (b) Sketch of sections of the BZ at given $k_z$ for systems with generalized $C_4$ and $C_6$ symmetry, respectively.
 The square and the hexagon denote the section of the BZ.
 When the indicator is nontrivial,
 exceptional loops emerge in the three-dimensional BZ.
 Correspondingly, exceptional points emerge in the two-dimensional subspace as denoted by green and blue dots.
 Here, blue (green) dots represent the exceptional points with $\nu=1$ ($-1$).
 In these panels, shaded area are denoted by the region enclosed by the integration path in Figs.~\ref{fig:BZ_GRS}(a) and \ref{fig:BZ_GRS}(b).
 }
\label{fig:BZ_GRS_3d}
\end{figure}
In this appendix, we show the details of the merging of exceptional loops arising by generalized rotational symmetry in three dimensions.
Consider the case in which the discriminant indicator $\xi_{C_n, 3D}$ ($n=4$, $6$) takes the nontrivial value.
In this case, the exceptional loops emerge, meaning that the presence of the exceptional point in a two-dimensional subspace is specified by a given $k_z$ (see Fig.~\ref{fig:BZ_GRS_3d}).
We note that an odd number of exceptional points exist in the shaded area in the BZ due to the nontrivial value of the two-dimensional indicator for this subspace.
We also note that applying the generalized rotation, an exceptional point is mapped to another one with the opposite sign of the velocity $\nu$ (see Fig.~\ref{fig:BZ_GRS_3d}).

The merging of the exceptional loops observed in Fig.~\ref{fig:TOY_GFRS_3d}(d) can be understood by shifting $k_z$.
Because of the indicator taking a value of $\xi_{C_4,3D} = -1$, the exceptional points observed in Fig.~\ref{fig:BZ_GRS_3d}(a) should annihilate each other.
Such annihilation is allowed only at $(k_x,k_y)= (0,0)$ and $(\pi,\pi)$ in the presence of generalized $C_4$ symmetry.
In the following, we see the details.

\begin{figure}[b]
 \centering
\includegraphics[trim={0cm 0cm 0cm 0cm},width =\hsize]{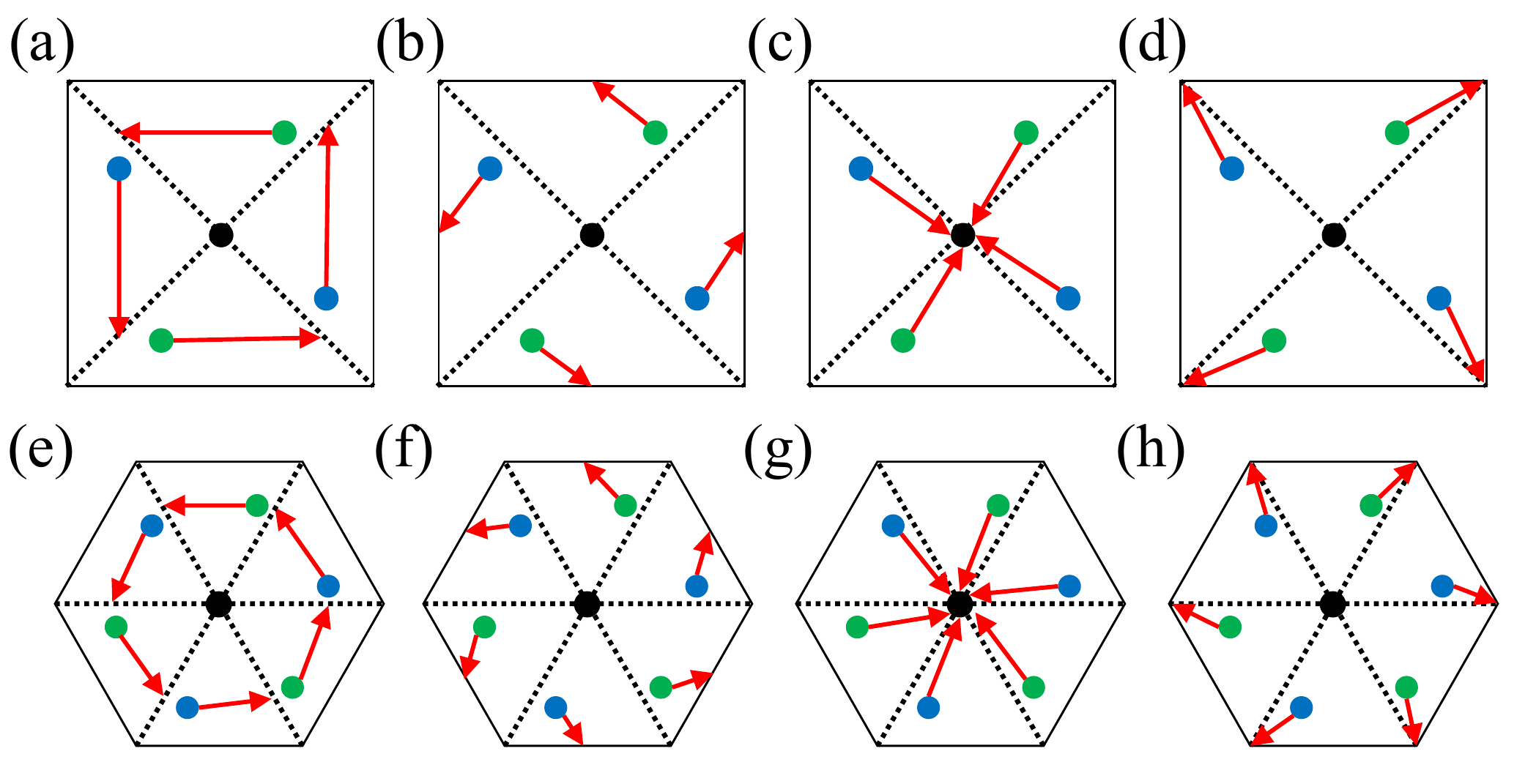}
\caption{
 (Color online)
 The path of the exceptional points in the BZ by shifting $k_z$ with generalized (a)$-$(d) $C_4$ and (e)$-$(h) $C_6$ symmetry.
 Blue (green) dots represent the exceptional points with $\nu=1$ ($-1$).
 The red arrow is the path of exceptional points by shifting $k_z$.
 }
\label{fig:BZ_PATH}
\end{figure}

Consider a proper $k_z$ where the exceptional points emerge [see Fig.~\ref{fig:BZ_GRS_3d}(a)].
Shifting $k_z$, we can see that these exceptional points in the shaded area move to the boundaries,
which is categorized into four cases [see Figs.~\ref{fig:BZ_PATH}(a)$-$\ref{fig:BZ_PATH}(d)].
In the case illustrated in Fig.~\ref{fig:BZ_PATH}(a), the exceptional point does not meet another one,
and thus,
pair annihilation does not occur.
In the case illustrated in Fig.~\ref{fig:BZ_PATH}(b),
the exceptional point meets another one but with the same vorticity, and thus,
pair annihilation does not occur.
In the case illustrated in Figs.~\ref{fig:BZ_PATH}(c) and \ref{fig:BZ_PATH}(d),
the exceptional point meets another one with opposite vorticities.
The above argument elucidates that two exceptional loops merge at the line specified by $(k_x,k_y)=(0,0)$ and $(\pi,\pi)$.

In a similar way, let us consider the case of generalized $C_6$ symmetry.
In this case, the indicator $\xi_{C_6,3D}=-1$ does not necessary mean the merging of exceptional loops.
Let us illustrate the details.
As is the case for the generalized $C_4$ symmetry, the movement of
the exceptional point for the pair annihilation is categorized into four cases in Figs.~\ref{fig:BZ_PATH}(e)-\ref{fig:BZ_PATH}(h).
In the case illustrated in Fig.~\ref{fig:BZ_PATH}(e), the exceptional point does not meet another one, and thus, pair annihilation does not occur.
In the case illustrated in Figs.~\ref{fig:BZ_PATH}(f) and \ref{fig:BZ_PATH}(g),
the exceptional point meets another one with the opposite vorticity.
Thus, pair annihilation occurs.
In the case illustrated in Fig.~\ref{fig:BZ_PATH}(h),
the exceptional point meets another one with the same vorticity,
and thus,
pair annihilation does not occur.
The above argument indicates that the indicator $\xi_{C_6,3D} = -1$ does not necessarily mean the merging of the exceptional points.


%\bibliographystyle{apsrev4-1}
%\bibliography{reference2}

\end{document}